\documentclass[graybox]{svmult}

\usepackage{type1cm}        
\usepackage{makeidx}         
\usepackage{graphicx}        
\usepackage{multicol}        
\usepackage[bottom]{footmisc}

\usepackage{newtxtext}       %
\usepackage{newtxmath}       

\usepackage{natbib}

\makeindex             

\begin{document}

\title*{Representing asymmetric relationships by h-plots. Discovering the archetypal patterns of cross-journal citation relationships}
\titlerunning{Asymmetric relationships by h-plots}
\author{Aleix Alcacer and Irene Epifanio}
\institute{Aleix Alcacer \at Jaume I University, Spain, \email{aalcacer@uji.es}
\and Irene Epifanio \at Jaume I University and ValgrAI, Spain \email{epifanio@uji.es}}
%
%
\maketitle

\abstract{This work approaches the multidimensional scaling problem from a novel angle. We introduce a scalable method based on the h-plot, which inherently accommodates asymmetric proximity data. Instead of embedding the objects themselves, the method embeds the variables that define the proximity to or from each object. It is straightforward to implement, and the quality of the resulting representation can be easily evaluated. The methodology is illustrated by visualizing the asymmetric relationships between the citing and cited profiles of journals on a common map. 
Two profiles that are far apart (or close together) in the h-plot, as measured by Euclidean distance, are different (or similar), respectively. This representation allows archetypoid analysis (ADA) to be calculated. ADA is used to find archetypal journals (or extreme cases). We can represent the dataset as convex combinations of these archetypal journals, making the results easy to interpret—even for non-experts. Comparisons with other methodologies are carried out, showing the good performance of our proposal. Code and data are available for reproducibility.}

\section{Introduction}
\label{sec:1}
Multidimensional scaling is a classical problem in many fields. Nevertheless, it remains a problem of interest, since in some fields such as bioinformatics, features are unavailable \citep[Ch.~18]{HTF09}, and only proximity information between pairs of objects is available. Therefore,  dimensionality reduction tools are fundamental in machine learning community.

In some fields such as image matching, text or web mining, or cognitive psychology, pairwise data are non-metric and the dissimilarity matrix does not satisfy the mathematical requirements of a metric function (reflexivity, definiteness, symmetry, triangle inequality). In these cases, non-metricity is not due to noisy measurements, but due to data being inherently non-metric. In these cases, we should not enforce metricity (for example, by adding a constant to the non-diagonal dissimilarities), as real information could be lost \citep{Duin}.

The great majority of well-established machine learning methods have been formulated for metric data only, and often the metricity violation is not taken into account, transforming the dissimilarities (for instance, by symmetrization) or simply omitting the negative eigenvalues as in classical scaling. 

However, there has been a resurgence of interest in the dimensionality reduction problem for asymmetric relationships, as evidenced by the surge of publications in this field over the last few years \citep{okada2024applied,olszewski2024asymmetric,olszewski2023asymmetric,di2025candecomp,bove2021methods}. An overview of different methods for asymmetric pairwise relationships can be found at \cite{bove2018methods}.

\cite{Epi2013} proposed a methodology based on the h-plot, which can take into account the non-metricity, including asymmetry and lack of reflexivity (positive selfdissimilarity). The idea is to embed the variables that define the proximity to or from each instance,  instead of embedding the instances themselves.  In the h-plot, the dissimilarity matrix is treated as a data matrix, representing two variables: the dissimilarity from element “j” to all other elements ($d_{j \cdot}$), and the dissimilarity from all other elements to “j” ($d_{\cdot j}$). In this mapping, the Euclidean distance between two variables represents the sample standard deviation of the difference between variables. If the two variables are similar, their difference—and consequently the standard deviation of that difference—will be small. 

Projecting into a Euclidean space allows us to use many statistical techniques, such as archetypoid analysis (ADA) \citep{Vinue15}, as made by \cite{vinue2017archetypoid}. ADA is an unsupervised \textcolor{black}{statistical} learning technique used to identify representative "extreme" cases (called archetypoids) within a dataset. These archetypoids are actual observations and are used to approximate all other data points as convex combinations of them.  ADA relies on Euclidean distances and convex combinations, so projecting the data into a Euclidean space (where these operations are well-defined and meaningful) enables the method to work properly. ADA summarizes data in a friendly human-interpretable way, as human beings interpret better the extremes.

\textcolor{black}{Clustering techniques could also be considered as an alternative to ADA, given that both approaches aim to uncover structure within data. However, important conceptual differences make their suitability context-dependent. Whereas clustering typically represents groups using central tendencies, ADA characterizes patterns through extreme yet representative observations, which can facilitate interpretation when data variation is better explained by boundary cases rather than central ones. Clustering is generally most effective when the data exhibit clearly separated groups, allowing cluster centers to meaningfully summarize each subgroup. In contrast, when group boundaries are diffuse or when the data form a continuum rather than distinct clusters, ADA may yield more informative results because its archetypoids capture the extremes that define the range of variation. Moreover, ADA provides the proportion or mixture weights with which each observation relates to each archetypoid, offering a nuanced description of how individual cases combine these extreme prototypes. A didactic example comparing both techniques is presented by \cite{IsmaelTFM}, providing an illustrative demonstration of their respective behaviors and interpretative advantages.}

The methodology is illustrated step-by-step by a data base about cross-journal citations of Statistical journals. In Sec. \ref{metode} the methodologies, h-plot and ADA, are introduced. \textcolor{black}{The data set is presented in Sec. \ref{datos}.} Its application is developed in Sec. \ref{res}. Finally, conclusions are provided in Sec. \ref{conc}.

For reproducibility, code and data are available at {\url https://epifanio.uji.es/RESEARCH/asyada.zip}.

\section{Methodology} \label{metode}
\subsection{H-plot}
H-plots were introduced by \cite{CorstenGabriel} for visualizing the relationships between multiple variables in a data matrix $\mathbf{X}$. For computing the h-plot in two-dimensions, the eigendecomposition of matrix $\mathbf{S}$, the variance–
covariance matrix of $\mathbf{X}$ is calculated. $\mathbf{S}$ is  always positive semidefinite, therefore their eigenvalues are always positive. Let $\lambda_1$ and $\lambda_2$ be 
the two largest eigenvalues and $\mathbf{q}_1$ and $\mathbf{q}_2$ be their corresponding unit eigenvectors. The h-plot in two-dimensions is: \[
\mathbf{H}_2 = \left( \sqrt{\lambda_1} \mathbf{q}_1,\ \sqrt{\lambda_2} \mathbf{q}_2 \right).
\]

The Euclidean distance between the rows $\mathbf{h}_i$ and $\mathbf{h}_j$ is approximately equal to the sample standard deviation of the difference between variables ``$i$'' and ``$j$'' \citep{Seber}. It would be equal for the full matrix  $
\mathbf{H}$. So, when the two largest eigenvalues account for the majority of the variance, the matrix $\mathbf{H}_2$ serves as an effective low-dimensional representation. The following goodness-of-fit measure proposed by \cite{CorstenGabriel} 
 is used for assessing h-plotting in two dimensions, where a value close to 1 indicates a better fit:

 \[
\frac{\lambda_1^2 + \lambda_2^2}{\sum_j \lambda_j^2}
\]

 Let $\mathbf{\Delta}$ be an $n \times n$ dissimilarity matrix  with elements $\delta_{ij}$, that represent the dissimilarity from object ``$i$'' to object ``$j$''. When $\mathbf{\Delta}$ is asymmetrical, let us consider $\mathbf{D}$ = $\left[ \mathbf{\Delta} \,\middle|\, \mathbf{\Delta}' \right]$, which is an $n \times 2n$ matrix composed by the combination by columns of $\mathbf{\Delta}$ and $\mathbf{\Delta}'$, where $'$ denotes the transposition. If we consider $\mathbf{D}$ as a data matrix, the first $n$ columns contain the variables of the kind $d_{\cdot j}$, i.e. the dissimilarity from other objects to ``j''; while, the second block of columns contains the variables of the kind $d_{j \cdot}$, i.e. the dissimilarity from ``j'' to other objects. After applying h-plotting, $\mathbf{H}_2$ has $2n$ rows. The first $n$ rows approximate  $d_{\cdot j}$ profiles \textcolor{black}{or descriptions}, while rows from $n + 1$ to $2n$ approximate $d_{j \cdot}$ profiles.

For multidimensional scaling, h-plotting is applied to $\mathbf{D}$ that means that if $d_{\cdot j}$ and $d_{\cdot i}$ profiles 
are similar, they will be represented close to each other; the same applies analogously for $d_{j\cdot}$ and $d_{i\cdot}$ profiles. Furthermore, $d_{\cdot j}$ and $d_{i\cdot}$ profiles are mapped within a unified representation, enabling direct comparison. When a $d_{\cdot j}$ profile is positioned close to a $d_{i \cdot}$ profile, it indicates similarity between them. Thus, by computing the Euclidean distance between $d_{\cdot j}$ and $d_{j \cdot}$ profiles within the representation, we can identify objects that are more or less asymmetric — that is, those whose $d_{\cdot j}$ and $d_{j \cdot}$ profiles align closely or differ significantly, respectively. 

In summary, the goal of the h-plot is not to \textcolor{black}{exactly} preserve interpoint dissimilarities\textcolor{black}{, nor to maintain the rank order of those dissimilarities.} Instead, it focuses on maintaining the relationships among dissimilarity variables, i.e. the profiles themselves. \textcolor{black}{As shown in Example 1 of \cite{Epi2013} there is a case in which the dissimilarity between $i$ and $j$ is small (the smallest distance is between MA and VL), yet this is not reflected in the h-plot projection, where the distance between MA and VL is only the second smallest. That is,  the ranking of dissimilarities is not preserved.}



\textcolor{black}{Note also that, although it would clearly be possible to represent the variables
$d_{i\cdot}$ and $d_{\cdot j}$ separately, doing so would result in the loss of
information arising from not considering them together. In particular, we would
not be able to determine whether a given {cited} profile of one journal is
similar to a {citing} profile of another. Nor would it be possible to
analyze the asymmetry of a given journal, since the {cited} and
{citing} profiles of the same journal would not be represented in a common
space. It should also be borne in mind that the axes need not have the same
interpretation in separate h-plot projections of {cited} and {citing}
profiles, and therefore different spaces cannot be meaningfully combined.}

\subsubsection{Advantages of h-plots}
Besides the easy interpretation of h-plots results, since all profiles are mapped in a unique representation, h-plots offer several advantages: (a) They provide an explicit solution based on eigenvectors, eliminating the issue of local minima common in other methods. Moreover, the solution remains consistent even when the roles of objects and individuals are reversed.
(b) H-plots are invariant under linear transformations of the dissimilarity scale, meaning the visual configuration remains unchanged even if the scale is altered \citep{Epi2013}.
(c) They are computationally efficient and can handle very large matrices \citep{schwarz2019scalable}, i.e it is scalable.
(d) Their goodness-of-fit is straightforward to evaluate; (e) in
case that some observations were outliers, robust h-plots could be computed as proposed by \cite{daigle1992robust}.

\subsection{Archetypoid analysis (ADA)}
Archetypoid analysis is based on the idea that data can be represented as a mixture of a set of archetypal patterns that belong to the dataset (see \cite{alcacer2025survey} for a survey in this field). Archetypal profiles in statistics carry the same intuitive meaning as in everyday life—they represent extreme or pure types—and are easier interpretable than central points, as humans tend to better understand contrasting or opposing components \citep{Thurau12}.  

Let us review ADA in the multivariate case.  Let ${\bf X}$ = (${\bf x}_1$, ..., ${\bf x}_n$) be an $n \times m$ data matrix with $n$ observations and $m$ variables.

In ADA, three matrices are defined to approximate the data structure through convex combinations of representative observations:

\begin{enumerate}
\item The $k$ archetypoids, denoted by $\mathbf{z}_j$, are the rows of a $k \times m$ matrix $\mathbf{Z}$, where each $\mathbf{z}_j$ corresponds to a specific, observed data point from the original data matrix $\mathbf{X}$.

\item An $n \times k$ matrix $\boldsymbol{\alpha} = (\alpha_{ij})$, which contains the mixture coefficients used to approximate each observation $\mathbf{x}_i$ as a convex combination of the archetypoids:
\[
\hat{\mathbf{x}}_i = \sum_{j=1}^k \alpha_{ij} \mathbf{z}_j.
\]

\item A $k \times n$ matrix $\boldsymbol{\beta} = (\beta_{jl})$, composed of binary coefficients that determine the selection of archetypoids from the data. Each archetypoid $\mathbf{z}_j$ is defined as:
\[
\mathbf{z}_j = \sum_{l=1}^n \beta_{jl} \mathbf{x}_l,
\]
where $\beta_{jl} \in \{0,1\}$ ensures that each archetypoid is a specific observation from $\mathbf{X}$.

\end{enumerate}

To estimate these matrices, the objective is to minimize the Residual Sum of Squares (RSS):

\begin{equation} \label{RSSar}
RSS = \displaystyle  \sum_{i=1}^n \| \mathbf{x}_i - \sum_{j=1}^k \alpha_{ij} \mathbf{z}_j\|^2 = \sum_{i=1}^n \| \mathbf{x}_i - \sum_{j=1}^k \alpha_{ij} \sum_{l=1}^n \beta_{jl} \mathbf{x}_l\|^2{,}
\end{equation}
where $|\cdot|$ denotes the Euclidean norm.

This optimization problem is subject to the following constraints:

\begin{enumerate}
\item Convexity constraint on the mixture coefficients:
$\displaystyle \sum_{j=1}^k \alpha_{ij} = 1$ with $\alpha_{ij} \geq 0$ {for} $i=1,\ldots,n$
\item Binary selection constraint for archetypoids:
$\displaystyle \sum_{l=1}^n \beta_{jl} = 1$ with $\beta_{jl} \in \{0,1\}$ and $j=1,\ldots,k$.

\end{enumerate}

These constraints ensure that each observation is approximated as a convex combination of actual data points, thereby enhancing interpretability and empirical validity.

The mixed-integer optimization problem underlying ADA is addressed using the two-phase algorithm proposed by \cite{Vinue15}. The first phase, known as BUILD, initializes the solution by selecting a set of candidate archetypoids. The second phase, referred to as SWAP, iteratively refines this selection by replacing elements from the initial set with unselected observations whenever such exchanges yield a reduction in the residual sum of squares (RSS). For our implementation, we utilize the R \citep{R} implementation developed by \cite{EpiIbSi17}. \textcolor{black}{Scalability of ADA algorithm was achieved through sampling by \cite{Vinue21}. Furthermore, the algorithm was also parallelized by \cite{Vinue21}, so more time can be saved.}

To determine the appropriate number of archetypoids, we apply the elbow criterion, a method previously employed in papers such as \cite{cutler1994archetypal, Eugster2009, Vinue15}. This approach involves plotting the RSS against the number of archetypoids and identifying the point at which the rate of decrease in RSS sharply changes—commonly referred to as the elbow.

\subsection{H-plot + ADA}
As in \cite{vinue2017archetypoid}, for obtaining archetypoids from h-plotting asymmetric relationships, let us note that each object ``j '' has two profiles: one representing $d_{j \cdot}$ and another approximating $d_{\cdot j}$. Both of them are represented in the same configuration by the two dimensional vectors  $\mathbf{h}_j$ and $\mathbf{h}_{j+n}$, with $j$ = 1, ..., $n$. Therefore,  we apply ADA to
an $n \times 4$ matrix ${\bf X}$ made up of the combination of the representation of the both blocks of h-plot profiles, i.e. the row $j$ of ${\bf X}$ is constituted by $\mathbf{h}_j$ and $\mathbf{h}_{j+n}$.

ADA has been applied with h-plotting projection in several previous works with good results. \cite{vinue2017archetypoid} used it with asymmetrical dissimilarities in a football team performance analysis; \cite{epifanio2023archetypal} applied it in the study of foot shapes, \cite{math9070771} in a shoe size recommendation system and \cite{doi:10.1080/00031305.2018.1545700} in several data sets with missing data. Furthermore, \cite{Vinue15} used it with non-metric but symmetric dissimilarities of 3D binary images of the trunk of Spanish women for apparel design. Moreover, \cite{Vinue15} carried out a simulation study with non-metric but symmetric dissimilarities to  determine which multidimensional scaling (MDS) method is more appropriate for recovering the archetypoids after the MDS projection. H-plot was compared with other six MDS methods (Classical
(Metric) Multidimensional Scaling, Kruskal’s Non-metric Multidimensional Scaling, Sammon’s Non-Linear Mapping, Isomap \citep{Tenenbaum}, Locally Linear Embedding \citep{RoweisLLE} and  Diffusion Map \citep{Coifman}), and  demonstrated a better ability to recover original archetypoids compared to the other MDS methods.

\textcolor{black}{As regards scalability, as both h-plot and ADA are scalable, its sequential combination is scalable.}

\textcolor{black}{Let us detail their computational complexity. For h-plot, first we compute the covariance matrix, which is mainly matrix multiplication, i.e.  $\mathcal{O}(n^3)$, and then eigenvectors are computed. The complexity of computing eigenvectors depends on the method and matrix. Although there are specialized algorithms that can be run even with \(O(1)\) \citep{sun2020memory}, in the worst case the computation can be done with standard algorithms in polynomial time \(O(n^{3})\). As regards ADA, the computational complexity depends largely on  the method used for estimating  $\boldsymbol{\alpha}$. If  the non-negative least squares (NNLS) \citep{Lawson74,cutler1994archetypal} is used, its complexity is \(O(n k ^{3})\)  \citep{alcacer2025survey}. Hence, the overall computational complexity is the product of the cost of enumerating all possible archetypoid combinations and the cost of the matrix $\boldsymbol{\alpha}$ update procedure, i.e.  \(O( \binom{n}{k} n k ^{3})\).
}



\section{\textcolor{black}{Data base}} \label{datos}
\cite{Epi2014} applied h-plot in the analysis of asymmetrical citation relationships with 25 journals in 2011 JCR edition (Journal Citation Reports by Clarivate) and from the subject
category “Statistics \& Probability". Here, we consider the first 25 journals by impact factor of the same category in 2025 JCR edition that provides 2024 data. Table \ref{revis} lists the name of the  journals together with its number in alphabetical order. \textcolor{black}{As previously noted, hplot + ADA is scalable and, therefore, can be applied to large databases without limitation. Nevertheless, for didactic and illustrative purposes, we chose to use a small database of only 25 observations in order to facilitate a more concise discussion of the results. This approach has also been adopted in other recent works, such as that by \cite{okada2024applied}.}

\begin{table}[ht]
\centering
\caption{List of Selected Journals} \label{revis}
\begin{tabular}{rp{0.9\linewidth}}
\hline
\textbf{\#} & \textbf{Journal Name} \\
\hline
1  & ANNALS OF PROBABILITY \\
2  & ANNALS OF STATISTICS \\
3  & ANNUAL REVIEW OF STATISTICS AND ITS APPLICATION \\
4  & BAYESIAN ANALYSIS \\
5  & BIOMETRIKA \\
6  & CHEMOMETRICS AND INTELLIGENT LABORATORY SYSTEMS \\
7  & ECONOMETRICA \\
8  & ECONOMETRICS AND STATISTICS \\
9  & ECONOMETRICS JOURNAL \\
10 & FUZZY SETS AND SYSTEMS \\
11 & IEEE-ACM TRANSACTIONS ON COMPUTATIONAL BIOLOGY AND BIOINFORMATICS \\
12 & JOURNAL OF STATISTICAL SOFTWARE \\
13 & JOURNAL OF THE AMERICAN STATISTICAL ASSOCIATION \\
14 & JOURNAL OF THE ROYAL STATISTICAL SOCIETY SERIES B-STATISTICAL METHODOLOGY \\
15 & LAW, PROBABILITY \& RISK \\
16 & MULTIVARIATE BEHAVIORAL RESEARCH \\
17 & PROBABILISTIC ENGINEERING MECHANICS \\
18 & QUALITY TECHNOLOGY AND QUANTITATIVE MANAGEMENT \\
19 & SPATIAL STATISTICS \\
20 & STATISTICAL ANALYSIS AND DATA MINING \\
21 & STATISTICAL SCIENCE \\
22 & STATISTICS SURVEYS \\
23 & STOCHASTIC ENVIRONMENTAL RESEARCH AND RISK ASSESSMENT \\
24 & TECHNOMETRICS \\
25 & WILEY INTERDISCIPLINARY REVIEWS-COMPUTATIONAL STATISTICS \\
\hline
\end{tabular}
\end{table}

Cross-journal citations are clearly asymmetrical, since journal 'i' does not necessarily cite journal 'j' to the same extent that 'j' cites 'i'. Likewise, the triangle inequality does not necessarily hold, and journals cite themselves (lack of reflexivity).

Instead of the raw inter-citation frequencies, we consider normalized frequencies by the relatedness factor \citep{pudovkin2002algorithmic}, to adjust for differences in journal size \citep{klavans2006identifying}. The journal relatedness of journal $i$ to journal $j$, denoted as $R_{i \to j}$, is defined by \cite{pudovkin2002algorithmic} as:

\[
R_{i \to j} = \frac{H_{i \to j} \cdot 10^6}{\text{Pap}_j \cdot \text{Ref}_i}
\]

where $H_{i \to j}$ is the number of citations in the current year from journal $i$ to journal $j$ (to papers published in $j$ across all years), $\text{Pap}_j$ is the number of papers published in journal $j$ in the current year, and $\text{Ref}_i$ is the number of references cited in journal $i$ in the current year.

Due to limitations in data availability, we restricted our analysis to citable articles published in 2022 and 2023. Accordingly, we considered only those citations made in 2024 that refer to these articles.

Given that our objective is pattern detection, adjusting the scale of the configuration (either expanding or contracting it) may provide more informative insights \citep{Seber}. For this reason, instead of using the relatedness measure values, we use their ranks (where the highest relatedness values receive the smallest rank) as done in \cite{Epi2013,Epi2014}. This form of relativization is also common in plant biology, where asymmetric measures are particularly effective for analyzing community data \citep{Podani}. In cases of tied values, we assign them the average rank among the ties. Those entries that are not defined because there are no cites, received the value one plus the maximum rank obtained (171 in our case). If similarities were necessary, dissimilarities could be converted into similarities by subtracting 171 from them.

\section{Results} \label{res}
\begin{figure}[ht]
\includegraphics[scale=.65]{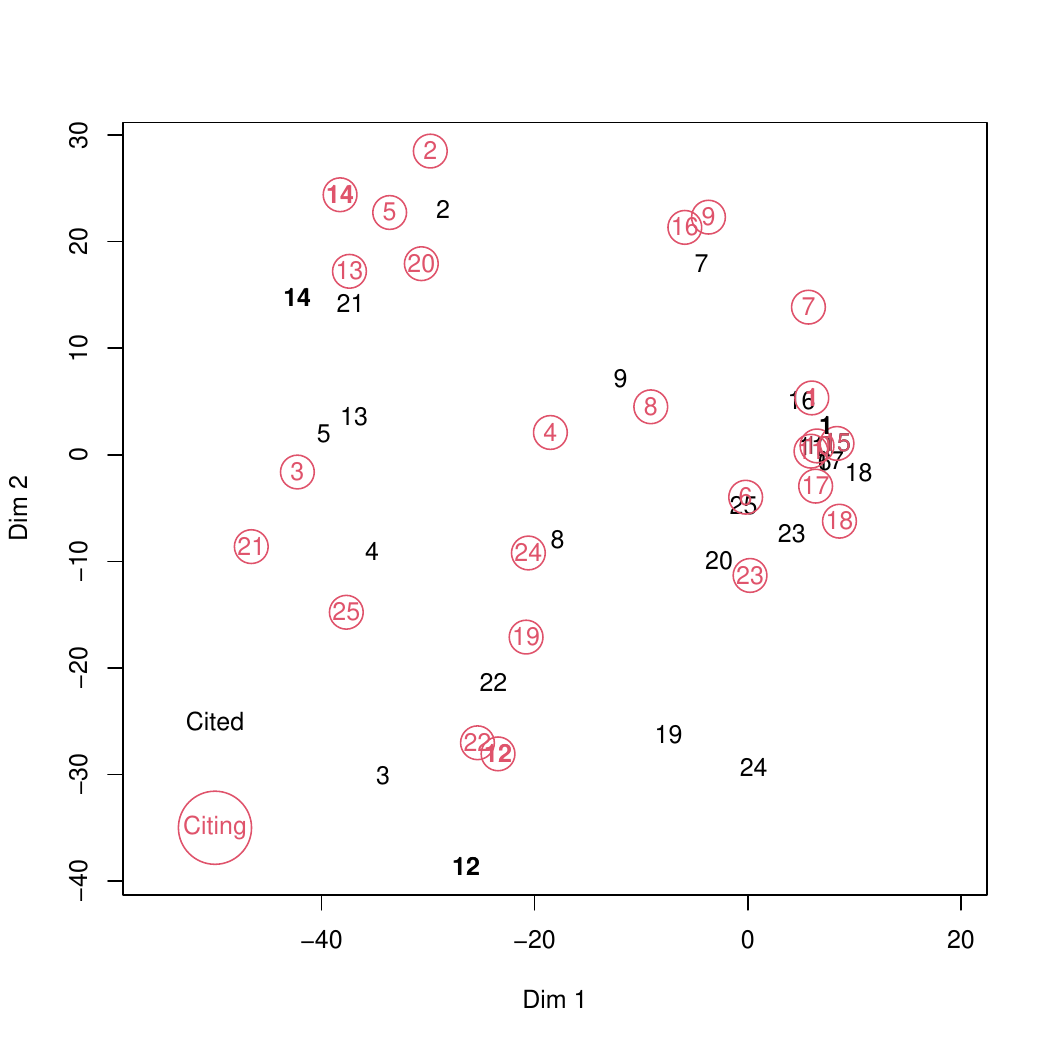}
%
%
\caption{H-plot mapping of 25 statistics journals. Table \ref{revis} provides the
codes for the numbers. \textcolor{black}{The bold font indicates the archetypoids for $k$ = 3.}}
\label{fig:hplot}       
\end{figure}

Fig. \ref{fig:hplot} displays the h-plot mapping of the journals.  The goodness-of-fit of the h-plot in two dimensions is 74.3\%, indicating a good representation. The positions of the cited journals are shown in black, while the citing positions are marked with red circles.

Let us begin by examining the citing profiles. At top left, there is a \textcolor{black}{set} of journals devoted to general theory
and methodology (2, 5, 13, 14, 20). At central left, we find a \textcolor{black}{set}  of review journals  (3, 21, 25). Two nearby journals are 12 and 22, which are both open access journals, one of them about surveys and the other about software. In the central part, there are journals halfway between theory and applications (4, 8, 19, 24). At the top right corner, there is a \textcolor{black}{set}  of social science and econometrics journals (7, 9, 16). Finally, at central right there are  journals in specific fields, such as applied in particular areas (6, 11, 17, 18, 23), probability (1, 15) and fuzzy sets (10). 

However, the cited profiles are not so grouped. Although, there are some \textcolor{black}{groups}, such as, in the top left (2, 5,13, 14, 21), corresponding to general theory
and methodology and review journals, and another \textcolor{black}{set}, mostly corresponding to journals in particular areas (1, 6, 10, 11, 15 16, 17, 18, 23) and STATISTICAL ANALYSIS AND DATA MINING (20), a journal halfway between theory and applications with a computational statistics profile as well as WILEY INTERDISCIPLINARY REVIEWS-COMPUTATIONAL STATISTICS (25), a review journal that is also specialized in reviewing in computational statistics.

Note that these two journals (20 and 25), together with ANNUAL REVIEW OF STATISTICS AND ITS APPLICATION (3), a journal that not only review methodological advances, but also the computational tools that allow for their implementation, are the most asymmetrical journals; i.e. they cite mostly certain journals, but are cited by other \textcolor{black}{kinds of} journals. For example, STATISTICAL ANALYSIS AND DATA MINING (20) cites mostly journals 13, 14 and 5; but, it is cited mostly by journals outside Statistics category.

On the other hand, the most symmetrical journals, i.e. those that cite and are cited in a similar way are: LAW, PROBABILITY \& RISK (15), FUZZY SETS AND SYSTEMS (10) and IEEE-ACM TRANSACTIONS ON COMPUTATIONAL BIOLOGY AND BIOINFORMATICS (11). They are journals in very specific fields.        

\subsection{ADA} ADA has been applied to the \textcolor{black}{h-plot} representation for discovering patterns and summarizing data. \textcolor{black}{We opted to employ ADA rather than clustering because Fig. \ref{fig:hplot} shows, even upon visual inspection, that the data do not exhibit well-separated groups; thus, ADA constitutes a more informative approach in this context. Furthermore, we computed the Average Silhouette Score, which is a global measure of clustering quality and observed that the highest value, 0.46, occurred at $k$ = 3. Since values below 0.5 indicate a weak cluster structure \citep{Kaufman90}, this finding corroborates the impression conveyed by Fig. \ref{fig:hplot}.}

\begin{figure}[ht]
\includegraphics[scale=.65]{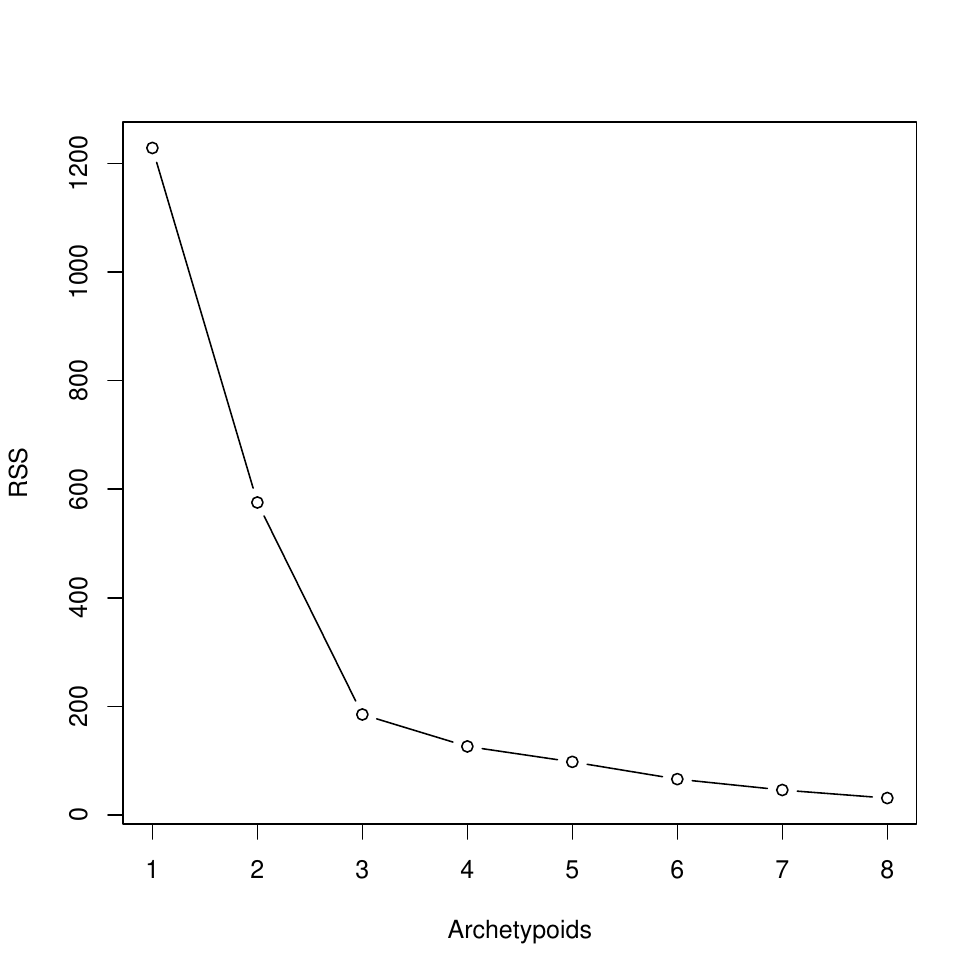}
%
%
\caption{Screeplot of ADA for the \textcolor{black}{h-plot} of 25 statistics journals.}
\label{elbow}       
\end{figure}

\begin{figure}[ht]
\includegraphics[scale=.65]{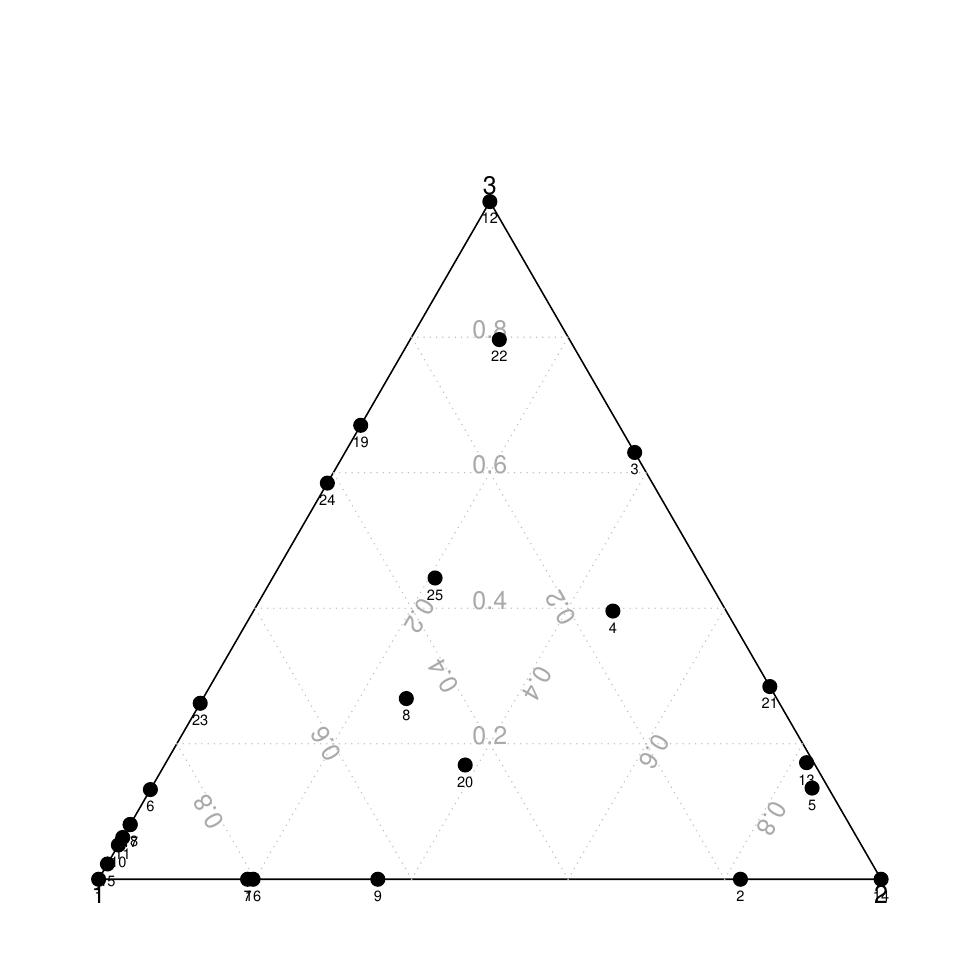}
%
%
\caption{Ternary plot of ADA for the \textcolor{black}{h-plot} of 25 statistics journals.}
\label{ternary}       
\end{figure}

Fig. \ref{elbow} shows the screeplot. There is an elbow at $k$ = 3. Fig. \ref{ternary} represents the $\alpha$ values for $k$ = 3, where the archetypoids are: ANNALS OF PROBABILITY (1),             JOURNAL OF THE ROYAL STATISTICAL SOCIETY SERIES B-STATISTICAL METHODOLOGY (14) and JOURNAL OF STATISTICAL SOFTWARE (12), i.e. a specific journal, a theoretical and methodological journal and a general applied computational statistical journal, respectively. With this representation, for example, ECONOMETRICS AND STATISTICS (8) is described as a mixture of 47\% archetyopid 1, plus 26\% arcehtypoid 2 and 27\% archetypoid 3. This fact reflects the scope of this journal that publishes papers in econometrics, statistical methodology and computational aspects. Analogously, ANNUAL REVIEW OF STATISTICS AND ITS APPLICATION (3) is approximated as 37\% archetypoid 2 plus 63\% archetypoid 3. This also fits the scope of the journal, which  reviews methodological advances and the computational tools.

Although we have focused on $k$ = 3 results, depending on the level of detail we are aiming for, we might consider more or fewer archetypoids.   With $k$ = 1, the archeypoid is the  medoid and corresponds with ECONOMETRICS AND STATISTICS (8), i.e. a eclectic journal. With $k$ = 2, the archetypoids are STOCHASTIC ENVIRONMENTAL RESEARCH AND RISK ASSESSMENT (23) and JOURNAL OF THE AMERICAN STATISTICAL ASSOCIATION (13), an applied and methodological journal, respectively. Although, archetypoids are not necessarily nested, in this data set happens. For $k$ = 4, the archetypoids JOURNAL OF THE ROYAL STATISTICAL SOCIETY SERIES B-STATISTICAL METHODOLOGY (14) and JOURNAL OF STATISTICAL SOFTWARE (12) coincide with those obtained with $k$ = 3, there is also a specific journal LAW, PROBABILITY \& RISK (15) as archetypoid and a review journal appears as archetypoid, WILEY INTERDISCIPLINARY REVIEWS-COMPUTATIONAL STATISTICS (25). With $k$
 = 5, again the archetypoids JOURNAL OF THE ROYAL STATISTICAL SOCIETY SERIES B-STATISTICAL METHODOLOGY (14),  JOURNAL OF STATISTICAL SOFTWARE (12), LAW, PROBABILITY \& RISK are preserved, the review journal is maintained with STATISTICAL SCIENCE (21), and a new archetypoidal profile appears with  TECHNOMETRICS (24) and applied statistical journal, but in engineering fields.

\subsection{Comparison with other methodologies}
\textcolor{black}{We have considered two well-established methodologies for which R packages are available and which have been previously applied to the analysis of cross-reference activity between journals: (a) multidimensional unfolding analysis, used  for instance  by \cite{schneider2009mapping} and \cite{Epi2014}, and (b) network representation, used for instance by \cite{calero2012seed} and \cite{Epi2014}.
}

\subsubsection{\textcolor{black}{Multidimensional unfolding analysis}}
\textcolor{black}{In multidimensional unfolding, data are typically represented as dissimilarities between the elements of two sets, consisting of $n_1$ individuals and $n_2$ objects. The objective is to identify a common quantitative scale that enables a visual examination of the relationships between these two sets. Multidimensional unfolding can be regarded as a special case of MDS.}

\textcolor{black}{Let $\boldsymbol{\Delta}$ denote the observed dissimilarity matrix of dimension $n_1 \times n_2$, with elements $\delta_{ij}$ representing the observed dissimilarity between individual $i$ and object $j$. The goal is to estimate a joint configuration matrix $\mathbf{X}$ of dimension $(n_1 + n_2) \times p$ by minimizing the following stress function:
\[
\sum_{i=1}^{n_1} \sum_{j=1}^{n_2} \left( \delta_{ij} - d_{ij}(\mathbf{X}_1, \mathbf{X}_2) \right)^2,
\]
where the distance function is defined as
\[
d_{ij}(\mathbf{X}_1, \mathbf{X}_2) = \left( \sum_{s=1}^{p} (x_{1is} - x_{2js})^2 \right)^{1/2}.
\]}

\textcolor{black}{
The matrix $\mathbf{X}$ is partitioned into two submatrices: $\mathbf{X}_1$ of dimension $n_1 \times p$, representing the configuration of individuals, and $\mathbf{X}_2$ of dimension $n_2 \times p$, representing the configuration of objects. In the present application, the two sets coincide, such that $n_1 = n_2 = n = 25$, and $p$ = 2.
}

\begin{figure}[ht]
\includegraphics[scale=.3]{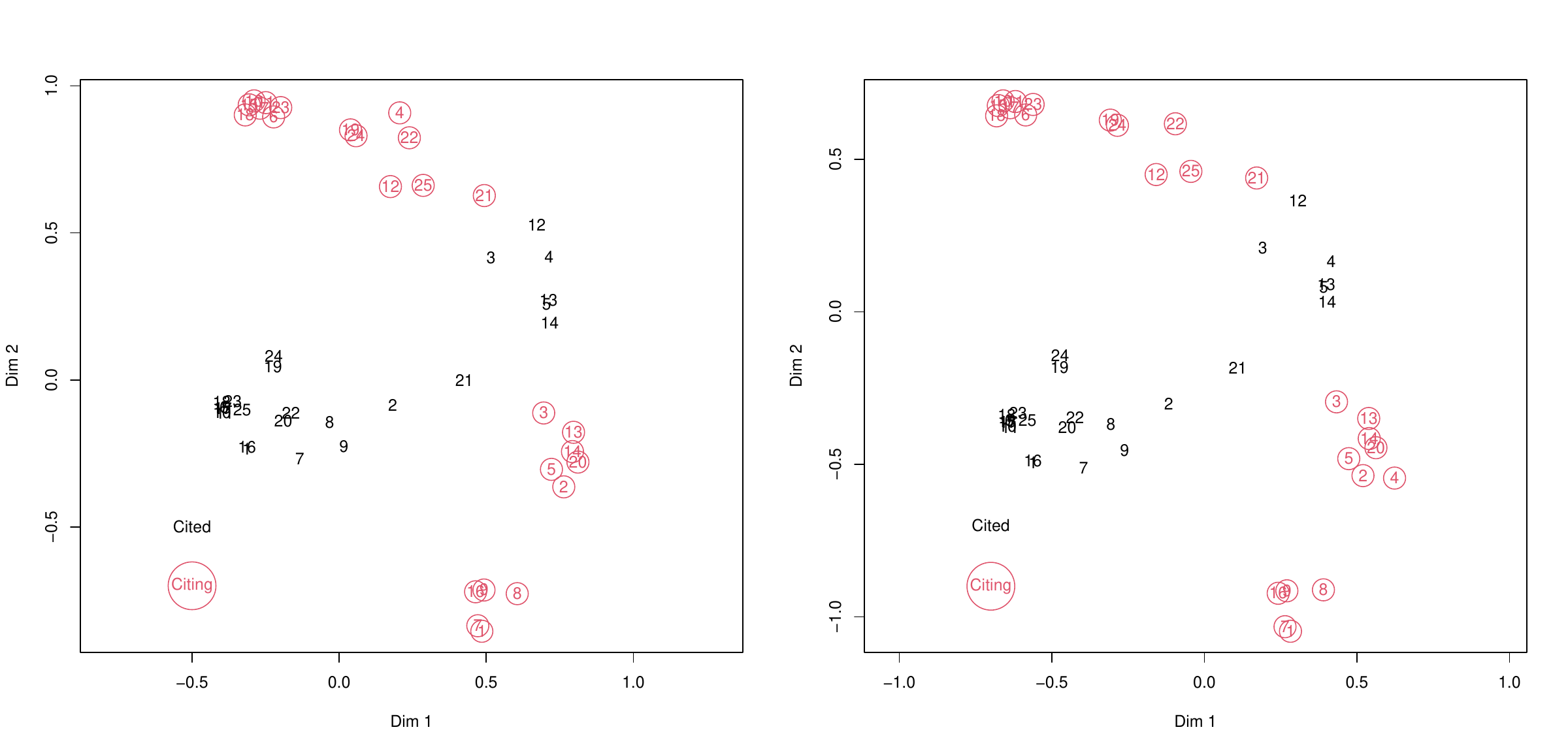}
%
%
\caption{Unfolding solutions of 25 statistics journals. Table \ref{revis} provides the
codes for the numbers.}
\label{fig:unfolding}       
\end{figure}

Multidimensional unfolding analysis results are shown in Fig. \ref{fig:unfolding}. This is computed by the R  package SMACOF \citep{de2009multidimensional,smacof}. On the left of, the best solution
is displayed when the roles of objects and individuals are
played by cited and citing profiles, respectively. These roles are exchanged, on the right. Configurations are similar in both situations\textcolor{black}{, although they are not exactly the same. If the global minimum were reached, the solution would remain unchanged regardless of the roles, but local minima tend to arise more frequently in low-dimensional solutions \citep{borg2017choice}, this is the reason why both roles were displayed}.

The cited positions are in the central part of the map, whereas the citing positions are grouped in two \textcolor{black}{sets} . On the one hand, the citing \textcolor{black}{set}  in the bottom right, in turn, is composed by two \textcolor{black}{groups}: a group of social (16) and econometrics journals (7, 8 and 9) and probability (1); and a \textcolor{black}{set}  with  journals about general theory
and methodology (2, 5, 13, 14) and with some applications (20), together with a review journal (3). The journal (4) is also in this group on the left of Fig. \ref{fig:unfolding}.  On the other hand, the \textcolor{black}{group}  in the top is divided into two \textcolor{black}{sets}: one more spread composed by review (21, 22, 25) and software (12) journals, together with journals halfway between theory and applications (19, 24); while, the other \textcolor{black}{set}  is tighter and is composed by journals of specific fields (6, 10, 11, 15, 17, 18, 23). These citing profiles are far apart from their corresponding cited profiles, although they are similar. For example, LAW, PROBABILITY \& RISK (15) is the most cited and (self)citing journal, and  the same happens with FUZZY SETS AND SYSTEMS (10), with 20\% or more of self-cites. More examples of incoherencies found in the unfolding mapping are: journals ANNALS OF PROBABILITY (1) and MULTIVARIATE BEHAVIORAL RESEARCH (16) are very close, when these two journals are quite different. Although the citing profile of ANNUAL REVIEW OF STATISTICS AND ITS APPLICATION (3) is similar to other review journals, such as STATISTICAL SCIENCE (21), they are apart, while they are near in h-plot solution.

\subsubsection{\textcolor{black}{Network representation}}

\begin{figure}[ht]
\includegraphics[scale=.65]{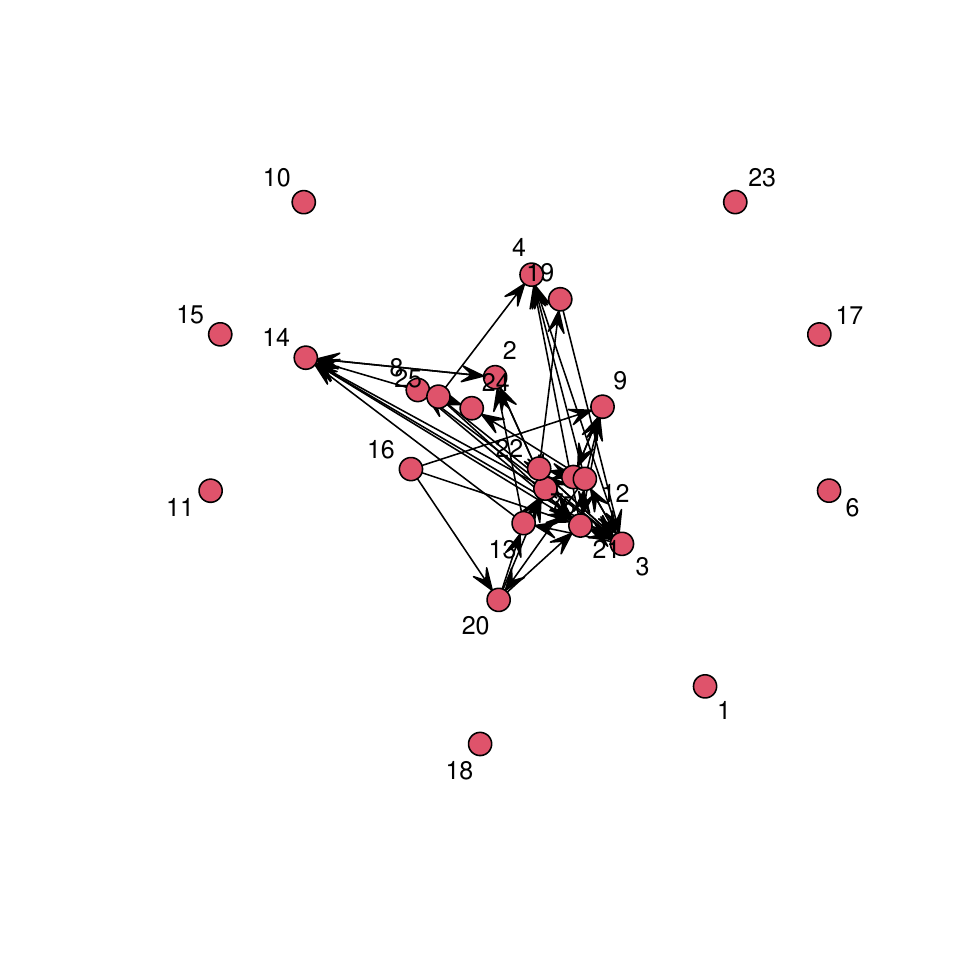}
%
%
\caption{Network representation of 25 statistics journals. Table \ref{revis} provides the
codes for the numbers.}
\label{fig:net}       
\end{figure}

The dissimilarity matrix can be converted into a similarity matrix and seen as a directed and
weighted graph and  represented via social
analysis visualization techniques, 
 as in  \cite{calero2012seed}. Fig. \ref{fig:net} shows a network representation with 100 as threshold in citation links with a spring embedder as layout obtained with R package sna \citep{butts2008social}. The threshold is used because the adjacency matrix is relatively dense, making the network representation hard to read \citep{ghoniem2005readability}. Even with the threshold and with the small size of our data set, the visualization is not easy to read, except for the isolated journals (1, 6, 10, 11, 15, 17, 18, 23), which corresponds with journals in specific fields. Furthermore, representing only the links we lose the information about how close the profile journals are.

\section{Conclusions} \label{conc}
This work demonstrates the potential of the h-plot as a practical and effective method \textcolor{black}{for representing asymmetric dissimilarities}. This is illustrated by mapping and analyzing asymmetric relationships between journals in terms of both citing and being cited. The h-plot can simultaneously represent both roles. The h-plot yields more informative results than unfolding \textcolor{black}{ and network representation}. Several advantages of using h-plots have been also highlighted. \textcolor{black}{Specifically, the advantages are: intuitive interpretability via a unified Euclidean representation; an explicit eigenvector-based analytical solution that avoids local minima; scale invariance under linear transformations; high computational efficiency for large matrices; and a straightforward evaluation of goodness of fit.}

Furthermore, archetypal profiles have been found by archetypoid analysis, which has allowed us to discover more easily the data structure.  \textcolor{black}{We identified three archetypal journals: a subject-specific journal, a theoretical and methodological journal, and a general applied computational statistics journal. For larger values of $k$, additional structural details emerge. In particular, when $k$ = 4, a review journal is identified as an additional archetypoid.}

\textcolor{black}{Although in this case we have employed ADA due to the absence of a clearly separated group structure in our data, h-plots allow the application of clustering techniques, to assist in uncovering the underlying data structure.}

\textcolor{black}{When the presence of outlying observations is suspected, robust h-plots may be computed following the approach proposed by \cite{daigle1992robust}. Analogously, robust ADA can be computed as presented by \cite{Moliner2018a}.}

Several directions for future research include representing symbolic dissimilarities by h-plots. \textcolor{black}{ Firstly,} in some cases, the exact dissimilarities are unknown, but information such as their range \citep{Groenen}  or a histogram \citep{groenen2006multidimensional} is available. For example, interval dissimilarities could be h-plotting using the ideas by \cite{d2021principal}. \textcolor{black}{ Secondly,}  another direction could be \textcolor{black}{performing} a modification of biarchetypal analysis \citep{alcacer2024biarchetype}  for finding archetypal profiles \textcolor{black}{directly from the dissimilarity matrix, by considering the same parameters for rows and columns}.   \textcolor{black}{Thirdly, h-plotting can be extended to handle}  two-mode three-way dissimilarities (object $\times$ object $\times$ source) \citep{okada1997asymmetric} by stacking the matrices of object $\times$ object  side by side \textcolor{black}{as developed by \citep{alcacer2025multidimensional}}.  \textcolor{black}{Fourth and finally, a new line could be considering supervised multidimensional scaling, i.e. taking into account  class labels or outcome variables, and defining a supervised h-plot through supervised principal component analysis \citep{barshan2011supervised}}.

\section*{Supplementary information}

 Code and data are available as supplementary information at {\url https://epifanio.uji.es/RESEARCH/asyada.zip}.

\begin{acknowledgement}
This work was partially supported by the Spanish Ministry of
Science and Innovation PID2022-141699NB-I00 and PID2020-118763GA-I00 and Generalitat Valenciana
CIPROM/2023/66.

\end{acknowledgement}

\bibliographystyle{apalike}
\bibliography{hplot.bib}
\end{document}